# Driving upside-down in a circular track

*Fernando F. Dall'Agnol, Lucas B. de Morais, Marcelo D. Alloy*

*Universidade Federal de Santa Catarina-Centro de Ciências Exatas e Educação-Blumenau. Rua João Pessoa 2514-Blumenau-SC, 89036-004, Brazil.*

## Abstract

In this article, we point out an interesting solution for the dynamics of a racecar in a banked circular track with banking angle well over 90º. We call this track configuration an Inverted Track, at which a racecar can drive partially upside-down. We show an experimental setup where we made a toy car to circulate upside-down held only by its friction to the track. We discuss the viability to perform the abovementioned stunt with a real racecar in terms of the velocities required, dimensions of the track and safety; provided a passionate motorsport related company to commission it. For most racecars, the aerodynamic down-force is significant and it is included in our analysis.

## 1. Introduction

A live experience of watching a two-metric-ton-stunt-car performing a loop is truly memorable. It is not like watching a rollercoaster where the carts are strongly attached to the rail. The former impresses more because the stunt car defies gravity while loose from the track. Unfortunately, these loopings are brief. If you blink, you may lose it. However, there is a solution for the dynamics of a stunt car, which can keep it upside-down in a circular track, not for a split second, but indefinitely. This solution is mathematically uncomplicated, yet, it has not been depicted in the literature; not even in movies or games.

There are massive analysis and academic exercises on the minimum velocity necessary to perform a loop, as well as for the dynamics on a banked curve. For example, driving on vertical Wall-of-death and Globes-of-death is an old and common stunt and they are very popular in Physics exercises [1,2]. However, apparently nobody have been curious enough to question how much more can the banking angle of a wall-of-death be increased beyond 90º and still sustain a car in circular motion. In this article, we show this relegated solution, under which a stunt car can be held (fairly) upside-down in a circular track at very high banking angles (~150º).

Some stunt enthusiasts have taken seriously the possibility of an F1 car to drive upside-down in a straight inverted lane [3,4]. This could be accomplished due to the aerodynamics of the F1 car that pushes it against the track. This stunt becomes more viable if we combine aerodynamic and friction forces on a circular inverted track, as we will discuss in detail.

## 2. The inverted circular track

Fig. 1(a) shows an ordinary circular track. As every driver should know, there is a maximum velocity in this track, over which a car skids out of the road. In the track in Fig .1(b), the maximum velocity is severely lessened because the banking angle is negative and the normal force pushes the car outward. Contrarily, if the track is banked inward, as in Fig. 1(c), pilots can go a lot faster as they do in many speedways [5]. In this case, there may be a minimum velocity (depending on the friction coefficient), under which the car slides down the track. We are interested in analyzing a circular racetrack banked inwards, particularly in angles larger than 90º, as in Fig. 1(d). In this case, the vector normal to the track points downward. In this article, we call tracks with banking angles larger than 90º as *inverted tracks*.

Solutions for the dynamics in Fig. 1(a), (b) and (c) are abundant. Most textbooks depict this cases, including the case for $\theta=90º$ [6,7]. On the other hand, it is safe to assert that the solution for the inverted track of Fig. 1(d) is very difficult to find. Possibly, it was never published.

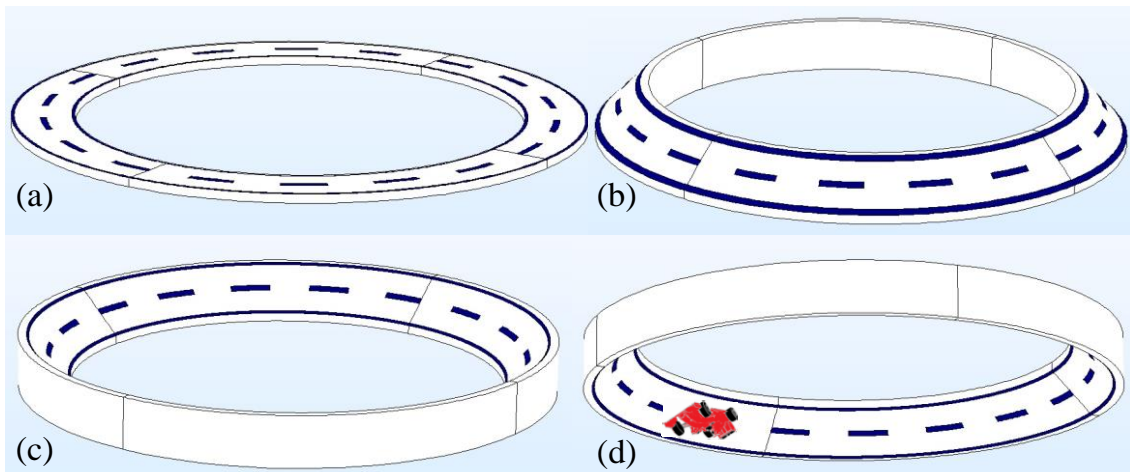

Fig. 1. Illustration of few representative tracks for four different banking angles. In (a) there is a simple circular track with no inclination. In (b) the outward bank diminishes the maximum velocity in this track. Cars in (c) can have a max and a min velocity, out of which it will skid out or slide in the track. In (d), there is a minimum velocity over which a stunt car can stay on the track, which may appear counterintuitive.

## 3. Solution for maximum and minimum velocities

Hereon, variables in bold represent vectors, whereas, the same variables in italic represent the norm of the corresponding vector. For example, **Fr** is the friction vector and *Fr* represents its norm.

Figure 2 indicates the forces that acts on a racecar; the weight $\mathbf{W}=(0,-mg)$, the normal $\mathbf{N}=(N_r, N_z)$ and the friction $\mathbf{Fr}=(Fr_r, Fr_z)$. The vector sum of all three must result in the centripetal force $\mathbf{Fc}=(-mv^2/R, 0)$, shown in dashed green.

Let's assume that the stunt car circulates counterclockwise and let $\mathbf{s}=(\cos\theta, \sin\theta)$ be a unit vector parallel to the lane pointing left to right from the perspective of the driver as indicated in Fig. 2(a). The **Fr** can point either in the direction of **s** or −**s** depending on the velocity *v*. There is a velocity $v_0=\sqrt{(gR\tan\theta)}$ where $\mathbf{Fr}=\mathbf{0}$. In this case, the racecar needs no help from the friction force to remain in circular motion [7,8]. If $v<v_0$, the car tends to slide to the left, but it is held by the friction force that points to the right in the **s** direction, as indicated in Fig. 2(a). There is a limit $v_{min}$, under which the friction cannot hold the car. At this limit, $\mathbf{Fr}(v_{min})=\mu N \mathbf{s}$. On the other hand, if $v>v_0$ the car tends to skid to the right, but it is held by the friction that points in the direction −**s**, as in Fig. 2(b). In this case, there is a maximum velocity $v_{max}$, over which the friction cannot hold the car. At this limit, $\mathbf{Fr}(v_{max})=-\mu N \mathbf{s}$. This analysis assumed $\theta$ in the first quadrant, for a typical banked track as in Fig. 1(c). Nevertheless, it is valid for any $\theta$ as follows:

- If $-90°<\theta<0$ (as in Fig. 1b), **Fr** points in the direction −**s** no matter the value of *v*. The car always tend to slide outward. In this case, the racecar skids off the track over a certain maximum velocity. At $v_{max}$, $\mathbf{Fr}(v_{max})=-\mu N \mathbf{s}$.

- If $90°<\theta<180°$ (inverted track /Fig. 1d), **Fr** has to point in direction **s**, as it is the only force in the system that can balance the weight. In this case, the car must have a minimum velocity to prevent it from falling off the track. At this limit, $\mathbf{Fr}(v_{min})=\mu N\mathbf{s}$.

- There is no solution for Eq.(1) for $-180°<\theta<-90°$.

In summary, we have the following conditions valid for all angles:

$$\mathbf{Fr}(v_{max}) = -\mu N\mathbf{s} = -\mu N(\cos\theta, \sin\theta), \tag{1}$$

$$\mathbf{Fr}(v_{min}) = \mu N\mathbf{s} = \mu N(\cos\theta, \sin\theta). \tag{2}$$

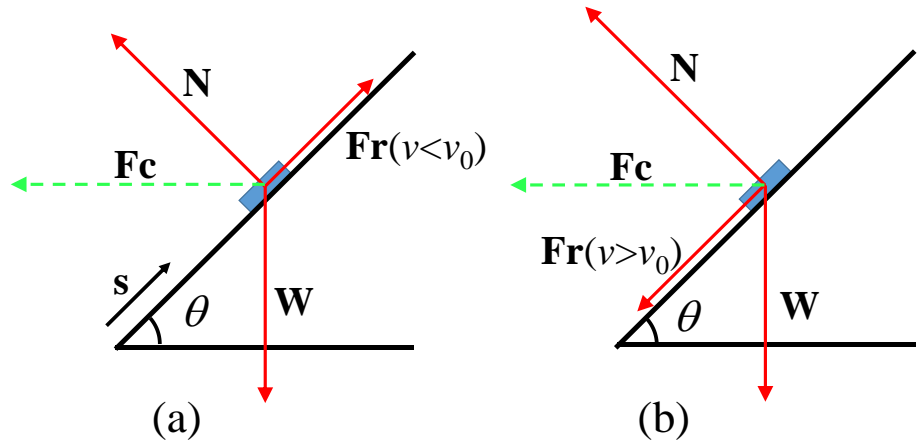

Fig. 2. Diagram of forces on a car in a banked curve. In (a), the $v$ is low therefore the car tends to slide to the left, so the friction force points to the right. In (b) the $v$ is high; the car tend to skid to the right, so the friction points to the left.

For a race car to stay on track the resultant force must be the **Fc**:

$$\mathbf{W}+\mathbf{N}+\mathbf{Fr}=\mathbf{Fc}, \tag{3}$$

$$\Rightarrow (0,-mg)+(N_r,N_z)+(Fr_r,Fr_z)=(-mv^2/r,0). \tag{4}$$

The components $N_r$ and $N_z$ can be expressed as (observe Fig. 2):

$$\begin{cases} N_r = N\cos(\theta + 90°) = -N\sin\theta \\ N_z = N\sin(\theta + 90°) = N\cos\theta \end{cases} \tag{5}$$

Replacing Eqs. (1) and (5) in (4) we have for $v_{max}$:

$$\begin{cases} \mu N\cos\theta + N\sin\theta = +\dfrac{mv_{max}^2}{R} \\ -\mu N\sin\theta + N\cos\theta = mg \end{cases} \tag{6}$$

Replacing Eqs. (2) and (5) in (4) we have for $v_{min}$:

$$\begin{cases} \mu N\cos\theta - N\sin\theta = -\dfrac{mv_{min}^2}{R} \\ \mu N\sin\theta + N\cos\theta = mg \end{cases} \tag{7}$$

The solutions for $v_{max}$ and $v_{min}$ in systems (6) and (7) are:

$$v_{max} = \sqrt{gR}\sqrt{\dfrac{\mu\cos\theta + \sin\theta}{\cos\theta - \mu\sin\theta}}, \tag{8}$$

$$v_{min} = \sqrt{gR}\sqrt{\dfrac{\sin\theta - \mu\cos\theta}{\mu\sin\theta + \cos\theta}}. \tag{9}$$

Fig. 3 shows the results for the $v_{min}$ and $v_{max}$ for a small and a large value of $\mu$. The shaded regions indicate the inverted track.

There are asymptotic angles where $v_{max}$ and $v_{min}$ tend to infinity. As $\theta \to \theta_{max}$, the $v_{max}$ must be very high for the car to skid out. As $\theta \to \theta_{min}$, the $v_{min}$ must tend to infinity for the car not to slide. Note that, in the inverted track ($\theta > 90°$), the larger the $\mu$ the lower the $v_{min}$ for a given $\theta$. For example, if $\theta = 100°$, then $v_{min}(\mu=0.3) = 332$ km/h and $v_{min}(\mu=1.3) = 119$ km/h. Equivalently, the larger the $\mu$ the more inclined the track can be for a given $v_{min}$. For example, if $\theta=135°$, then $v_{min}(1.3)=315$ km/h and the solution for $v_{min}(0.3)$ does not exist.

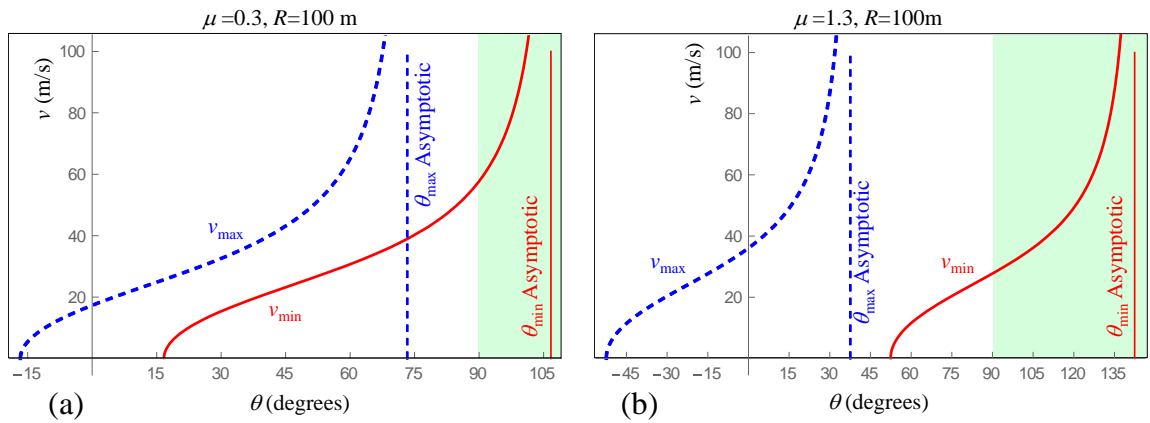

Fig. 3. Solutions for the max and min velocities as a function of the banking angle. In (a) the friction coefficient is low, consequently, there is a large range of angels where the min and max velocities can coexist. In (b), the friction coefficient is high. Notice that the minimum velocities necessary to drive inverted ($\theta > 90°$) are lower.

## 4. Solutions with aerodynamic force

In this section, we analyze the influence of the aerodynamic force $\mathbf{A}=(A_r,A_z)$ only for the minimum velocity $v_{min\_aero}$. Fig. 4 shows the forces acting on the car in this case.

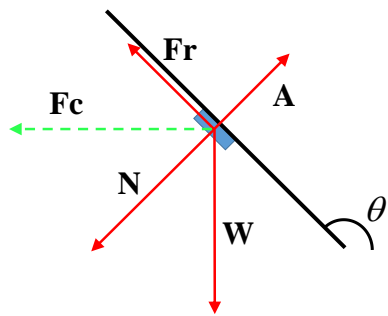

Fig. 4. Diagram of forces including the aerodynamic "downforce" $\mathbf{A}$, which points upward on the inverted track.

The norm of the aerodynamic force is modeled as being proportional to the square of the velocity [9],

$$A = \alpha v^2. \tag{10}$$

The vector **A** is then:

$$\mathbf{A} = \alpha v^2 (\cos(\theta - 90°), \sin(\theta - 90°)) = \alpha v^2 (\sin\theta, -\cos\theta). \tag{11}$$

With aerodynamics, Eq. (3) becomes

$$\mathbf{W} + \mathbf{N} + \mathbf{Fr} + \mathbf{A} = \mathbf{Fc}, \tag{12}$$

$$\Rightarrow \begin{cases} \mu N \cos\theta - N \sin\theta + \alpha v_{min\_aero}^2 \sin\theta = -\dfrac{m v_{min\_aero}^2}{R} \\ \mu N \sin\theta + N \cos\theta - \alpha v_{min\_aero}^2 \cos\theta = mg \end{cases}. \tag{13}$$

The solution of Eqs.(13) involves basic algebra, which we will not present here. The result for $v_{min\_aero}$ is:

$$v_{min\_aero} = \sqrt{gR} \sqrt{\dfrac{\sin\theta - \mu\cos\theta}{\mu\sin\theta + \cos\theta + \dfrac{\alpha\mu R}{2m}(2\cos^2\theta + \sin 2\theta \tan\theta)}}. \tag{14}$$

This solution for $v_{min\_aero}$ is the same as for Eq. (9) except for the third term in the denominator. This term is always positive for the inverted track (90<$\theta$<180), therefore, $v_{min\_aero} < v_{min}$, as expected.

Fig. 5 shows a comparison between the $v_{min}$ and $v_{min\_aero}$, assuming $\alpha = 10$ Ns$^2$m$^{-2}$ [9], $m = 1000$ kg [10], $\mu = 1.3$ [11]. To drive at the inclination of 135°, our results show that the car must have $v_{min} = 315$ km/h (88 m/s). However, the aerodynamic downforce drastically reduces the minimum velocity to $v_{min\_aero} = 118$ km/h (32.8 m/s). Eq. (14) predicts the minimum velocity to hold a car upside-down in a straight lane only by means of the aerodynamic force, by doing $R \to \infty$ and $\theta=180°$. In this case we have $v_{min\_aero}$=114 km/h (31.6 m/s).

The aerodynamic force makes it possible the for a stunt car to circulate in a track over 180°, as illustrated in Fig. 6. For example, assuming the parameters as in previous paragraph with $\theta=225°$ and $m$=760 kg, then Eq. (14) predicts $v_{min\_aero}$=50 m/s (180 km/h). Maybe one day this could be tried using unmanned vehicles.

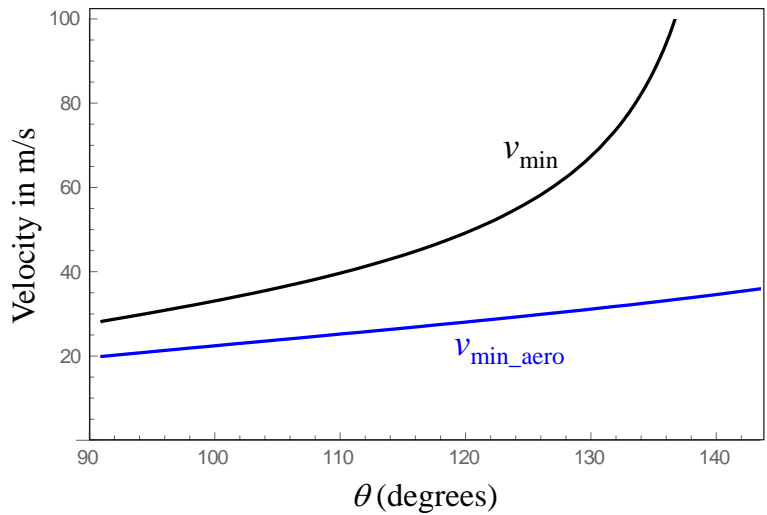

Fig. 5. Comparison between the minimum velocities to drive in an inverted track with and without the aid of aerodynamic forces.

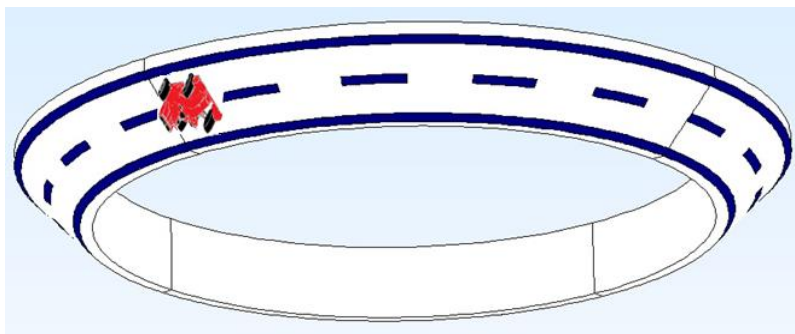

Fig. 6. Illustration of an F1 car driving at 225° of inclination. With sufficient aerodynamic downforce, it is possible to drive on tracks of any inclination.

## 5. Experiment

Notice that rotating a cylindrical track with a stunt car resting on the wall is equivalent to the car rotating on a motionless cylindrical track. In both cases, just the static friction is involved between the tires and the track. In our experiment, we take advantage of this equivalence because it is much easier to make the whole track plus the toy to spin than making a functional electric toy car to drive under an inverted track. In the former, it is not even necessary to build the whole track, just a segment large enough to fit the toy. Fig. 7(a) shows our system mounted on a bicycle wheel. The banking angle is $\theta=125°\pm1$. We coated the track segment in sandpaper and the car wheels where coated in soft rubber to get a large friction. The static friction coefficient we obtained was $\mu=1.10\pm0.06$. The average and standard deviation are from 15 measurements using a slanted plane technique [12]. The radius of the circumference is $R_{toy}=30.0\pm0.5$ cm from the axis of rotation to the center of mass. With this values, Eq.(9) predicts $v_{min}=3.6\pm0.3$ m/s.

Initially, a pair of magnets hold the toy in place as illustrated in Fig. 7(b). One of the magnets is glued underneath the toy and the other (the contermagnet) is placed behind the track. Then, the wheel is spun by unwinding a string from a spool attached to the spokes. A gentle pull of the string easily accelerate the toy to velocities greater than $v_{min}$, right after the first cycle. At a certain rotation, at about the third cycle, the countermagnet is centrifuged away, so the toy becomes held only by friction. The velocity eventually decays below $v_{min}$ and the toy falls off. Please watch the video in Ref. [13] showing our experiment. Notice, in Fig. 7(b), that we use a tube perpendicular to the track to house the countermagnet. This tube is important to guide the countermagnet straight outward. Otherwise, the countermagnet skids down the back of the platform, dragging the toy with it. Conveniently, the tube can accommodate a spacer to increase the distance between the magnet-countermagnet pair, so that the rotation speed needed to separate them can be adjusted. A third magnet (stopper) at the end of the tube holds the centrifuged countermagnet.

From the video, we can infer the velocity of the toy when it falls off the track. There are 138±1 frames in the last turn. Each turn takes 1s/240, so the velocity measured is $v_{min\_exp}$=3.28±0.06 m/s, which is exactly the value predicted using Eq. (9).

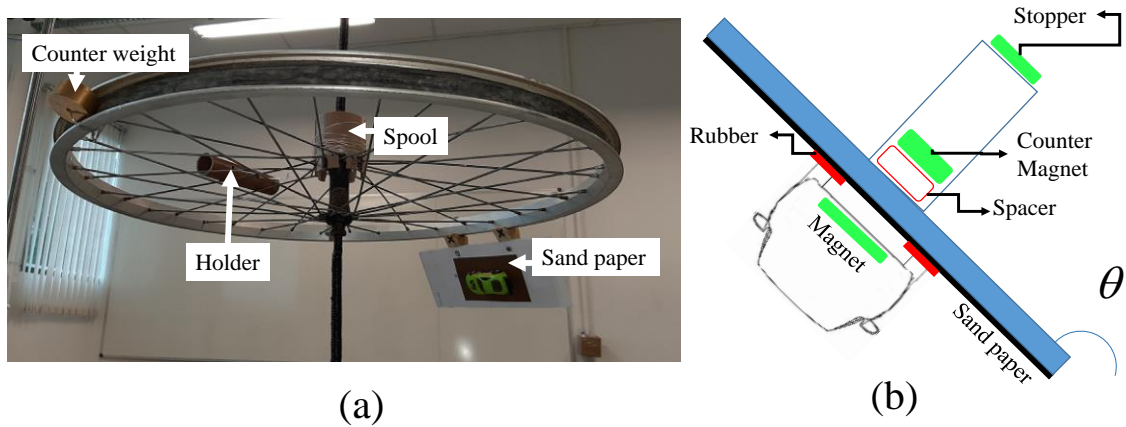

Fig.7. Experimental setup to hold a toy upside-down in circular motion only by friction. In (a) there is the photograph of our setup. In (b) the schematic representation of the car at its platform. While motionless, the magnet-countermagnet pair holds the car in position. When the spool is unreeled, the toy's velocity quickly surpasses $v_{min}$ and the countermagnet is centrifuged away, then, the toy is held only by friction. The toy eventually falls when the velocity decays below $v_{min}$.

## 6. Viability of making this stunt to happen for real.

The possibility to drive in an inverted circular track may raise great interest among car racers, stunt enthusiasts and motorsport-related companies. In the video from Ref. [4], the professional pilot Scott Mansell discusses this subject. In his video, Scott proposes a ~ 400 m straight upside-down lane, under which an F1 car could drive for ~10 s. This length does not include the track lengths necessary to take on and off the inverted track. We suggest a circular track instead, as illustrated in Fig. 8(a). We believe this concept is more advantageous for the following reasons:

- Since the circular track does not end, it allows the pilot to patiently transition from the upside-right to the upside-down positions little by little (See Fig. 8b).
- It should be safer because the car is permanently being centrifuged. So, if the velocity ever drops below $v_{min\_aero}$, the pilot won't just plummet head first. If something goes wrong, the car is expected to skids toward lower inclinations until it recovers the grip to the track.
- The pilot can stay upside-down for as long as he/she wants. Albeit not 180º upside-down, just ~135º upside down. Nevertheless, 135º should be sufficiently impressive.

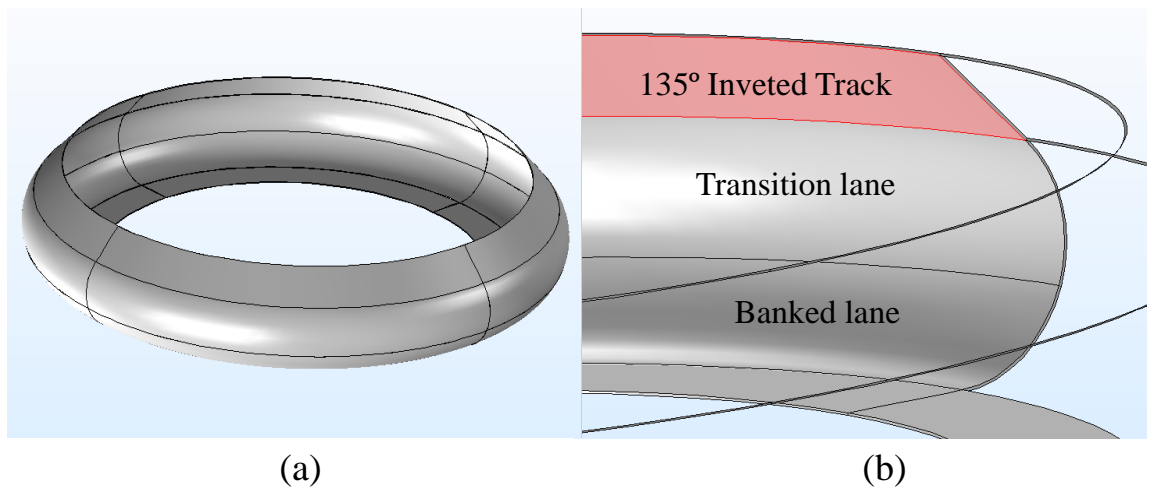

(a)         (b)

Fig. 8. Proposed track to drive an F1 car upside-down. In (a) there is the overall view of the speedway shaped as an arena (aspect ratio not to scale). In (b) there is a close-up view of the track's cross section, showing transition zones where the horizontal to inverted position can be achieved progressively.

## 7. Conclusions

We have presented a solution for a rather simple and interesting academic problem. This solution predicts the minimum velocity necessary to perform a counterintuitive stunt, in which a racecar can drive upside-down indefinitely in a circular inverted track. If the solution involves aerodynamic down-forces (as most racecars generates), the minimum velocity to stay upside-down decreases a lot. If an inverted track is ever built for this stunt, as some people seriously consider, we recommend the circular inverted track instead of a straight track. The circular track should be safer and more impressive, since the driver can stay upside-down for a long time.

## References

[1] https://www.youtube.com/watch?v=V0ZmzWc6YGI. Globe of Death video from 1955

[2] https://www.youtube.com/watch?v=BJ2f675LgMQ. Scary looking Wall of Death in India.

[3] T. Glossop, S. Jinks, R. Hopton, "Can an F1 car drive upside-down?" Phys. Spec. Top., **9**, 2010.

[4] https://www.youtube.com/watch?v=tFZ9loSAIPA&t=106s. Scott Mansell video.

[5] https://www.reference.com/sports-active-lifestyle/nascar-track-curves-steepest-bank-5aa11f283325685.

[6] https://www.youtube.com/watch?v=ZiaedxHszxg. Rotor at a French amusement park.

[7] D. Halliday, R. Resnick, J. Walker, *Fundamentals of physics*, (Wiley & Sons, New Jersey, 10th edition), v1, p. 135-138.

[8] http://www.batesville.k12.in.us/physics/phynet/mechanics/circular%20motion/banked_no_friction.htm.

[9] Joseph Katz, "Aerodynamics of race cars", Annu. Rev. Fluid Mech., **38**, 27-63, (2006).

[10] Formula One technical regulations 2017. Available at https://www.fia.com/file/54257/download/18380?token=5JfbyV2g.

[11] Database from http://www.mulsannescorner.com/data.html.

[12] V. B. Teffo, N. Naudé, J. S. Afr., Inst. Min. Metall. **113**(4) 351-356. 2013.

[13] https://youtu.be/7XE8Wu6dMgw. Video with our experimental setup and discussions.